\documentclass[12pt,preprint]{aastex}

\usepackage{amsmath, amsthm, amssymb}
\usepackage{natbib}
\usepackage{amssymb}
\usepackage{graphicx}
\usepackage{multirow}
\usepackage[normalem]{ulem}
\usepackage{color}
\newcommand {\llGRBs} {{\it ll-}GRBs}
\newcommand {\llGRB} {{\it ll-}GRB}
\title{Are low luminosity GRBs generated by relativistic jets?}
\author{\large Omer Bromberg$^1$, Ehud Nakar$^2$, Tsvi Piran$^1$\\
\footnotesize $^1$ Racah Institute of Physics, The Hebrew University, 91904 Jerusalem, Israel\\
\footnotesize $^2$ The Raymond and Berverly Sackler School of Physics and Astronomy,\\
\footnotesize Tel Aviv University, 69978 Tel Aviv, Israel}
\begin{abstract}
Low luminosity gamma-ray bursts (\llGRBs) constitute a sub-class of
gamma-ray bursts (GRBs) that plays a central role in the GRB-supernova connection.
While {\llGRBs} differ from  typical long GRBs (LGRBs)  in
many aspects, they also share some common features.  Therefore, the
question whether the gamma-ray emission of {\llGRBs} and
LGRBs has a common origin is of great interest.
Here we
address this question by testing whether {\llGRBs}, like
LGRBs according to the Collapsar model,
can be generated by relativistic jets that punch holes in the envelopes of
their progenitor stars.
The collapsar model predicts that the durations
of most observed bursts will be comparable to, or longer than, the
time it takes the jets to breakout of the star. We calculate the jet
breakout times of {\llGRBs} and compare them to the
observed durations. We find that there is a significant access of
{\llGRBs} with durations that are much shorter than the
jet breakout time and that these are inconsistent with the Collapsar
model. We conclude that the processes that dominate the gamma-ray
emission of {\llGRBs} and of LGRBs are most likely fundamentally different.
 \end{abstract}

\begin{document}

\section{Introduction}

%There  is a long line of evidence connecting long GRBs (LGRBs) to
%collapsing massive stars \citep[for recent reviewes
%see][]{Woosley06, Hjorth11}. Six LGRBs  were accompanied by
%spectroscopically confirmed broad-line
%Ic supernovae (SNe).
%The afterglows of about two dozens LGRBs showed "red bumps", a
%photometric evidence of underlying SNe. Last but not least, the
%identification of  LGRB host galaxies as highly star forming
%galaxies and the localization of the LGRBs in the most active star
%forming regions within  those galaxies
%\citep{Bloom02,Le Floc'h03,Christensen04,Fruchter06},
%({\it (Bloom et al. 2002a, Le Floc'h et al. 2003,
%Christensen et al. 2004, Fruchter et al. 2006)}),
%provides an
%indirect evidence for the connection of LGRBs with young massive
%stars and with stellar death.

% outlines a theoretical link between LGRBs and SNe.
According to the Collapsar model \citep{Paczynski98,MacFadyen99} the
core collapse of a massive star results in the formation of a
compact object, a black hole or a rapidly rotating neutron star. The
compact object ejects
%,  at ts center.  Following this collapse
a relativistic bipolar, baryon poor jet, along its rotation axis.  %and it is  accelerated to relativistic velocities.
The jet  punctures the surrounding stellar envelope and it emits the observed $\gamma$-rays
%when its energy is dissipated,
at a large distance from the star where the optical depth is small and the high energy photons can escape.
This model, that  is accepted as the standard model for
long GRBs (LGRBs),  explains naturally the association of
some LGRBs with SNe, and their general emergence in
star forming regions  \citep[see][for recent reviews]{Woosley06, Hjorth11}.

%%Somewhat surprisingly, a
A closer look at the members of the  spectroscopically confirmed
GRB/SN group (on which the GRB/SNe association hinges) reveals that four out of the six detected bursts:
GRB980425 (SN1998bw), GRB031203 (SN2003lw), GRB060218 (SN2006aj) and
GRB100316D (SN2010bh), are quite different  than ``normal"
LGRBs (see \S 2):  They are
%%much
less luminous;
%%they
have a smooth non-variable lightcurve,
and
%%their spectra
show no evidence for a high energy power-law tail.
%peak spectral energy is significantly lower than the
%typical peak energy of LGRBs.
Although only a handful of such bursts were
observed, the small observable volume set by their low luminosity
implies an event rate much
%%above
higher than the rate of LGRBs  pointing
towards Earth
\citep{Coward05,Cobb06,Pian06,Soderberg06,Liang07,Guetta07, Fan11}.
The unique characteristics of these low luminosity GRBs
(denoted hereafter {\llGRBs}) suggest that they may be generated by a totally
different process than most LGRBs.
As such it is of great interest to check whether %the gamma-ray emission of
{\llGRBs} %have indeed a different origin than that of LGRBs, or whether they
can arise from Collapsars.
%\textcolor{red}{\sout{To address this we address the question}}
Specifically  we ask the question:
can {\llGRBs} be generated
%, like ``standard"  LGRBs,
by relativistic jets that break out of their
progenitor stars.

To answer this question we study, in \S 3, (following Bromberg et al. 2011; hereafter B11)
the propagation of a relativistic jet in a stellar envelope.  We obtain the minimal conditions
required for the jet to break out of the star. %and form a successful GRB.
Specifically, we estimate the minimal time that the
central engine must power the jet for a successful crossing
of the star. Using this minimal breakout time we examine the expected duration distributions of LGRBs, SGRBs and
 {\llGRBs} (\S 4). We  discuss the implications of this distribution on the origin
of {\llGRBs} in \S 5 and we summarize our results  in \S 6.

 %The paper is structured as follow:

%We review, in \S 2,  the characteristics of
%{\llGRBs} and the expected properties of their jets.
%We summarize  the jet propagation model of B11, in \S 3,
%and obtain the expressions for the jet breakout time, and the minimal
%energy the jet must carry.
%We examine the distribution of the durations of {\llGRBs}
%and
%discuss the implications of this distribution on the origin
%of {\llGRBs}.
%We summarize the results and their implications in \S 6.

\section{The properties of {\llGRBs}}

{\llGRBs} are characterized by
%have distinct features that distinguish them from most
%LGRBs. The
isotropic equivalent luminosities, $10^{46}-10^{48}$ erg/s, that are
much lower than typical, $10^{51}-10^{53}$ ergs/s, emitted by LGRBs.
The durations range between $\sim10$ sec to an hour (in an extreme
case of GRB 060218), and the corresponding isotropic equivalent
energies of  $E_{\gamma}=10^{48}-$ a few times $10^{49}$ ergs, are
two to three orders of magnitude lower than those of typical LGRBs.
Apart from the low-luminosity which defines this sub-group,
{\llGRBs} have a softer  spectrum  with  typical peak energies
significantly below the average  of LGRBs and with no evidence of
high energy tail. Finally, {\llGRBs}'  lightcurves are smooth, each
containing only a single pulse. Most {\llGRBs} are accompanied by
%\textcolor{blue}{Another important property of {\llGRBs} is the association of all events with}
energetic broad line type Ic SNe with a strong radio emission.
Radiation models ascribe the radio emission to a mildly relativistic
shock moving ahead of the non-relativistic SN material
\citep[e.g.][]{Kulkarni98}. Rebrightening  episodes in the radio
emission are commonly associated with additional supply of energy
that refresh the shock, indicating  the presence of an internal
engine that can operate for long times \citep{Li99}. Finally, the
afterglow of some {\llGRBs} also show indications for late time
activity of such an engine \citep{Soderberg06}.

Due to their low luminosity, {\llGRBs} are detected only from  low
redshifts ($z\lesssim0.1$). With these redshifts, the four observed
{\llGRBs} imply an event rate of $230^{+490}_{-190}$
Gpc$^{-3}$yr$^{-1}$ \citep[][see also Coward 2005; Cobb 2006; Liang
et al. 2007; Guetta \& Della Valle 2007; Fan et al.
2011]{Soderberg06}, about 100-1000 times higher than the rate of
LGRBs pointing toward earth
\citep[][]{Coward05,Liang07,Guetta07,Wanderman10}.
\citet{Soderberg06} estimated the rate of broad line Ibc SNe to be
of the same order as the rate of {\llGRBs}, implying that {\llGRBs}
cannot be significantly beamed and that
 they could very well be isotropic. Using the overall ratio of the rates of broad line type Ib,c
SNe and {\llGRBs} we find that the  beaming factor of {\llGRBs} is $\lesssim 10$,  corresponding to
opening angles $\gtrsim30^\circ$.

The lack of bright, late time, radio emission from  {\llGRBs}
strongly constrain   the total energy of any relativistic outflow
involved in these events \citep{Waxman04,Soderberg04a,Soderberg06b}.
Additionally statistical arguments rule out the possibility that
{\llGRBs} are  regular LGRBs viewed at a large angle
\citep[e.g.][]{Daigne07}. Thus,  if {\llGRBs} are generated by
relativistic jets these jets must be weak and have a large opening
angle.
 %In the following sections we investigate the propagation of
%such a low luminosity wide jet within a Collapsar, and the implication on the source of {\llGRBs}.

\section{Jets Propagations in Stellar Envelopes}

We review, briefly,  the essential features of  jet propagation in a stellar envelope  (B11).
%that are relevant for this work.
Consider a cold relativistic
jet with a power $L_j$ and an initial opening angle $\theta_0$ that is
injected into a star.
As the jet propagates it pushes the stellar material in
front of it, leading to the formation of a double shock structure at
the jet's front, %which we refer to as
the jet's head. The pressure of the shocked material is much higher
than the pressure of the surrounding medium, thus matter that enters
the head is heated and pushed sideways forming a pressurized cocoon
surrounding the jet. The cocoon, in turns,  applies a pressure on
the jet and if the jet power is not too large it collimates the jet
into a cylindrical shape. The material in the collimated jet remains
relativistic and its Lorentz factor is
$\Gamma_j\simeq\theta_0^{-1}$. The jet's head  propagates, however,
at a much lower velocity and  it effectively dissipates all the
jet's energy into the cocoon. Thus, in order for the jet to
breakout, the engine must operate continually and supply power to
the jet practically until the jet's head reaches the surface, at
which stage the dissipation stops.

The  jet propagation depends on the stellar density profile.
%, which can generally be classified into two regimes.
Above the stellar core through a considerable fraction of the
envelope, where the jet spends most of its propagation time,  the
density can be approximated as a power law with an index
$1.5\lesssim\alpha\lesssim3$. %\citep{MatzMck99}.
%Outside this region the density profile steepens and it drops sharply towards the edge of the star.
For all the relevant parameter regime in {\llGRBs} and regular LGRBs
the jet's head is sub relativistic throughout this region. In this
non-relativistic limit the  head's velocity, $\beta_h\ll1$,
satisfies:
\begin{equation}\label{eq:bh}
\beta_h = \kappa\left(\frac{L_j}{t^2\rho c^5\theta_0^4}\right)^{1/5},
%\left[\frac{16}{3\pi}\varrho\zeta^{-1}\right]^{1/3}
\end{equation}
where $\rho$ is the density of the star at the position of the head, and $\kappa$ is a constant of
order unity that   depends on the power-law index of the density profile (B11).
As the head reaches the stellar edge, where the density drops sharply, it accelerates and for all practical purposes the jet can be considered as having escaped from the star.
The stellar radius, $R=c\int_0^{t_B}\beta_h dt$, is, therefore, the breakout radius of the jet from the star.
Using  eq. (\ref{eq:bh}) we obtain  the breakout time:
\begin{equation}\label{eq:tB}
t_B\simeq30~sec\cdot L_{47}^{-1/3}\theta_{10^\circ}^{4/3}R_{11}^{2/3}M_{15\odot}^{1/3},
\end{equation}
where $L_{47}$ is the jet's power in units of $10^{47}$ erg/s; $\theta_{10^\circ}$ is
the opening angle in units of $10^\circ$; $R_{11}$ is the stellar radius in units of $10^{11}$cm
and $M_{15\bigodot}$ is the stellar mass in units of 15 solar masses.
Here we use an average density profile of $\rho\propto r^{-2.5}$. The result changes only slightly
for different profiles.

As long as the jet propagates in the star, the head dissipates its
energy into the cocoon. If the engine shuts off before $t_B$, i.e.,
before the head reaches the surface, the head will stall. The energy
deposited by the jet into  the cocoon at a time $t$ is $L_j t$.
Since $\beta_h\ll1$, this implies that the minimal energy required
for the jet to cross the star is $E_{min}\simeq L_jt_B$, giving:
\begin{equation}\label{eq:E_min}
E_{min}\simeq3\times10^{48}~ergs\cdot L_{47}^{2/3}\theta_{10^\circ}^{4/3}R_{11}^{2/3}M_{15\odot}^{1/3}.
\end{equation}
%After breakout the head no longer dissipates energy.
%Therefore $E_{min}$ is also the total energy  deposited in the stellar envelope in a successful breakout.

The estimates of the breakout time and minimal energy depend
on the jet's properties  inside the star, which are not observed directly.
However we can relate them to the  observed properties of the GRB.
At late times, when the jet has evacuated a channel in  the stellar
envelope, its opening angle practically equals to the injection angle $\theta_0$.
It can be shown that this holds, in a non trivial way, also at earlier time,
just after the jet breaks out from the star.
%\textcolor{blue}{Thus measurements of the jet opening angle from jet breaks give
%a good estimate for $\theta_0$.
%It can be shown that at earlier times and possibly during the prompt emission, the
%opening angle cannot be larger than $\theta_0$ and may be somewhat smaller than that \citep{Lazzati09,B11}.
%Here we take the view that the angle is equal to $\theta_0$ and discuss the implications
%from a smaller opening angle at the conclusions.}
%Second, since
As there is no direct feedback between the jet that crosses the envelope and the central engine, the observed luminosity of the jet (after breakout) should be comparable to the jet's luminosity while it propagates %during the propagation phase
in the stellar envelope.
This allows us to estimate the jet breakout time %, given in eq. (\ref{eq:tB})
using the observed
GRB's prompt isotropic equivalent luminosity, $L_{\gamma}$ and the observed opening angle, $\theta$:
\begin{equation}\label{eq:tB_GRB}
t_B\simeq15~sec\cdot \eta^{1/3}L_{\gamma,50}^{-1/3}\theta_{10^\circ}^{2/3}R_{11}^{2/3}M_{15\odot}^{1/3},
%t_B\simeq10~sec\cdot L_{\gamma,50}^{-1/3}\eta_{-1}^{1/3}\theta_{10^\circ}^{2/3}R_{11}^{2/3}M_{15\odot}^{1/3},
\end{equation}
where $L_{\gamma}=\eta L_j \frac{2}{1-cos\theta_0}$\footnote{Note that $L_j$ is the luminosity of each one of
the two jets.}  and $\eta$ is the efficiency of
converting the jet power to radiation. Finally, similar reasonings,
namely the fact that the activity of the central engine is
determined by the stellar core whose initial radius is $\sim10^8$ cm
while the propagation of the jet takes place on a much larger scale
and is determined by the structure of the envelope which is only
weakly coupled to the core's mass \citep[e.g.][and references there
in]{Crowther07}, suggest that the duration that the central engine
operates, $t_{eng}$, should be independent of the jet breakout time.

\section{{\llGRBs} as Collapsars?}

The duration of the prompt emission, approximated by the proper
value of $T_{90}$, cannot be shorter than the time that the engine
is active  after the jet breakout. In most GRB models the two are
equal and  $T_{90} = t_{eng} -  t_B$. As discussed earlier,
within the Collapsar model  $t_{eng}$ and $t_B$ are uncorrelated.
This  implies that without fine tuning we expect that only a small
fractions of bursts should have $T_{90}\ll t_B$. Namely, it is
unreasonable that generically the engine operates just long enough
to let the jet break out of the star and then stops  right after
breakout. This is a direct implication of the Collapsar model and if
{\llGRBs} arise from Collapsars they should satisfy this condition.

To test the hypothesis that {\llGRBs} are Collapsars we examine
their duration distribution and compare it with the duration
distribution of regular {\it Swift}  LGRBs. Our sample contains the
four observed {\llGRBs} and  the  {\it Swift} LGRBs with measured
redshifts. We calculate the isotropic equivalent luminosity of the
{\it Swift} bursts by dividing the observed fluence in the BAT band
(15-150 keV) with the observed $T_{90}$ and correcting for redshift.
The result is multiplied by 3 to account for the total energy
radiated in all bands. We set $L_{\gamma}=2\cdot10^{48}$, the
 highest luminosity among the four confirmed
{\llGRBs}, as a threshold luminosity and consider any burst with a lower
luminosity to be a {\llGRB}. We find one additional burst
(GRB051109B) that matches the low luminosity criterion.
Interestingly, apart from fulfilling the luminosity criterion, the
light curve of this burst is also smooth and single peaked  and  no
strong emission is detected in the 100-350 keV band,  suggesting a
relatively low spectral peak like in other {\llGRBs}
\citep{Hullinger05}.  As there are no records of a SN search in the
error box of this burst during the time it could have been detected,
we cannot rule out the existence of an
associated SN.
%Consequently we consider  also this burst as {\llGRB}.
The bursts
with $L_{\gamma}>2\cdot10^{48}$ are considered as regular GRBs  and
are separated into LGRBs and SGRBs according to the standard
criterion of whether  $T_{90}$ in the observer frame is above or
below 2 sec.

For each burst with a given observed luminosity we calculate
the expected jet breakout time from a $15M_\odot$ star with a radius
of $10^{11}$ cm,  assuming an opening angle of $10^\circ$. For  the
four {\llGRBs} associated with SNe we use the mass estimates from
the associated SN (see table 1) and an opening angle of $30^\circ$.
Finely, to estimate the jet power we use a radiative efficiency
coefficient $\eta=1/2$. Changing the progenitor radius between
$5\cdot10^{10} - 5\cdot10^{11}$ cm and the radiative efficiency
between $0.1-1$ doesn't significantly change our results.

\begin{figure}
\includegraphics[width=7in]{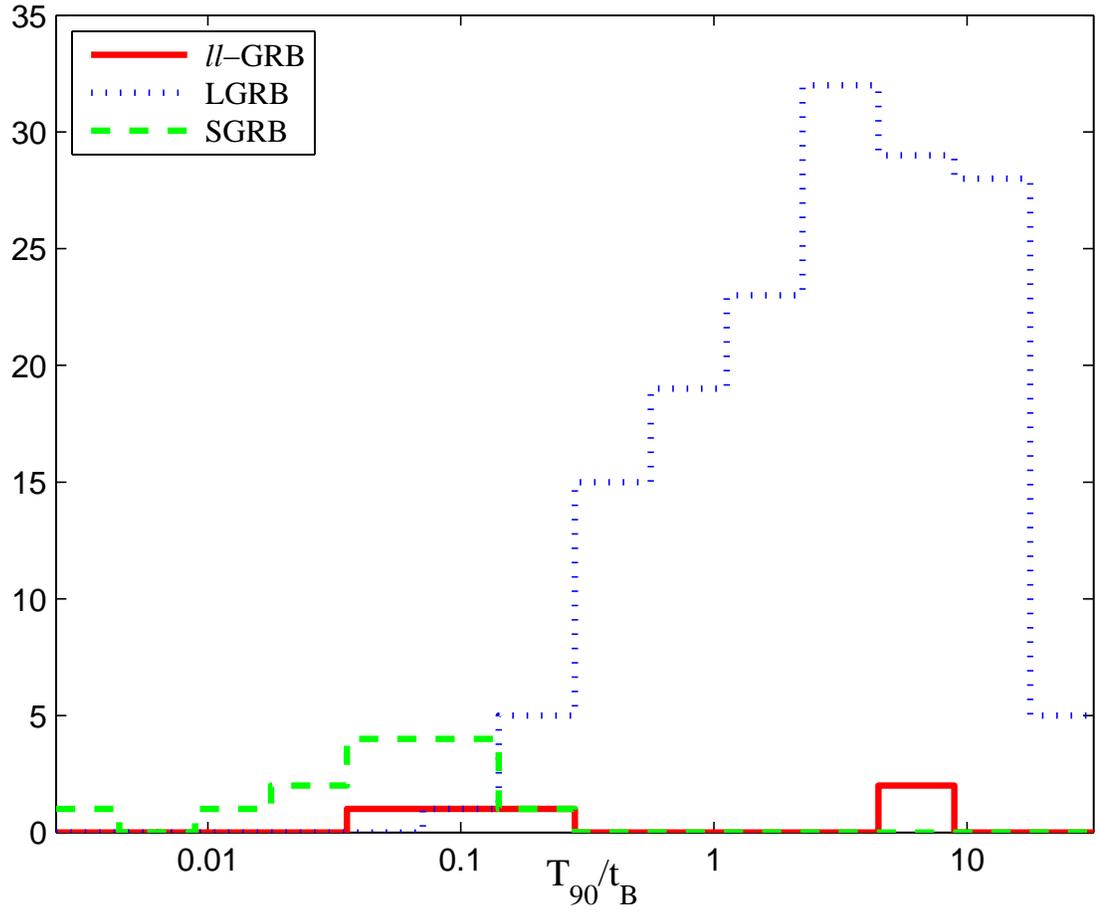}
  \caption{The distribution of $T_{90}/t_B$ for LGRBs,
  {\llGRBs}, and SGRBs.
 % Our samples contains the four known {\llGRBs} plus {\it Swift} LGRBs with redshifts.
 }

\label{fig.T90_T_tot}

\end{figure}

 Fig. \ref{fig.T90_T_tot} depicts the  distributions of
%$T_{90}/(T_{90}+t_B)$
$T_{90}/t_B$ of  {\llGRB}, SGRBs  and LGRB.  About $20\%$ of LGRBs
have $T_{90}<t_B$, in agreement with the expected small probability
of having  $t_{eng}\simeq t_B$ in a jet that successfully breaks
out.
%Consider now, the group of SGRBs. One can clearly see that
All SGRBs are concentrated at low values of
%$T_{90}/(T_{90}+t_B)$
$T_{90}/t_B$ with $T_{90} < 0.2 t_B$. This is a manifestation of
the well accepted concept that SGRBs cannot arise from a jet
breakout and cannot result from Collapsars.

Although there are two {\llGRBs} with $T_{90} > t_B$ the overall distribution of
 {\llGRB} differs  significantly from that of LGRBs and it  closer to  the distribution of SGRBs.
%Most {\llGRBs} have $T_{90}\ll t_B$. %, which is unexpected within the
%Collapsar model.
In particular,  3 out of 5 {\llGRBs}  have
$T_{90}<0.25t_B$, while  less than 4\% of  LGRBs are
in this range. Using the K-S test we
can estimate the chance that the observed duration distribution of
{\llGRBs} is taken from the LGRBs duration distribution. With such a few data points the
standard $\chi^2$ distribution doesn't give a good estimate for the
probability to get a given K-S distance. To remedy this  we use a Monte Carlo K-S to
estimate this probability.
% to obtain the measured K-S distance
%between the two commutative distributions.
%We preformed a K-S teat to evaluate the maximal distance between the
%cumulative distributions of the two populations.
%We normalized the distribution of the LGRBs to get a distribution function.
We randomly drew 5 events from the LGRBs distribution and obtain the
K-S distance between the simulated sample and the LGRBs distribution. We
repeated this process $10^5$ times and find that %more
less than $5\%$ of the randomly chosen events have larger K-S
distance than the {\llGRB} sample. %, implying that the probability that the sample of {\llGRBs} is taken from the LGRB populations is $<4\%$.
This suggests that the origin of {\llGRBs} is most likely
different than that of LGRBs. In particular the large fraction of
events with $T_{90}\ll t_B$ in the {\llGRB} sample disfavors the
successful jet break scenario.
% as a possible source for the high energy photons.
%A comparison of the {\llGRBs} to SGRBs gave a similar result where the chance that
%the two distributions are identical is $<4\%$.

%Note that

Unlike the LGRB sample that contains only {\it  Swift} bursts, the
{\llGRBs} sample includes bursts from three different detectors:
{\it  Swift}, INTEGRAL and BATSE,. This introduces selection effects
that are hard to quantify. Nevertheless, there  is no clear effect
against detection of short duration ($T_{90}\ll t_B$), soft GRBs, by
{\it Swift} or by the two other detectors. Since the discrepancy
between {\llGRBs} and LGRBs populations is dominated by such events,
we do not expect the heterogenic composition of {\llGRBs} sample to
affect our conclusion.

The Collapsar  model introduces two independent time scales,
$t_{eng}$ and $t_B$. 
This provides a  simple way to quantify the expected $T_{90}$
distribution for bursts with $T_{90}\ll t_B$ \citep{B11b}. Let
$p_{eng}(t_{eng})$ be  the probability density for $t_{eng}$. Let
$t_B$ be a typical jet breakout time.
The number of bursts per unit of
duration time, $T_{90}$, is :
\begin{equation}\label{eq.t90_dist}
{d N}/d T_{90}= p_{eng}(T_{90}+t_B)  .
\end{equation}
Assuming that $p_{eng}$ is smooth and does not vary on short
time scales in the vicinity of $t_b$, then for $T_{90}\ll t_B$, $d N
/d T_{90} \simeq p_{eng}(t_B) $,
which is a constant, independent of $T_{90}$.
Indeed, %this prediction of a flat
the observed
$T_{90}$ distribution of LGRBs,  satisfies this prediction,
providing an additional support to
the Collapsar model \citep[see][for a thorough discussion]{B11b}.

For a typical {\llGRB} luminosity and for the observed progenitor
stars that are implied by their SNe, the typical $t_B$ is of order of
$\gtrsim 200$ sec. 
There are two {\llGRB}   with $10<T_{90}\lesssim20$ sec. Thus for a flat
distribution, we would expect 
$\sim 20 $   with
$20\lesssim T_{90}\lesssim200$ sec, where only one 
was observed. 
The probability that those three events
are randomly selected from a flat distribution
between 0 and 200 sec 
is $<3\% $. This is in contrast to LGRBs where this distribution is flat \citep[see][]{B11b}
Even though the sample of the {\llGRBs} is very
small, the large fraction of bursts with $T_{90}\ll t_B$, combined
with the lack of accompanied bursts with $T_{90}\sim t_B$, implies
that it is highly unlikely that this distribution arises from
Collapsars.
While the statistical significance is lower, the reasoning leading
to this conclusion is
similar to the one that have lead to the realization that SGRBs
cannot be produced by Collapsars.

\begin{table}[h]
  \centering
  \caption{}\label{table_zhang}  
  \vspace{2mm}
  \begin{tabular}  {c c c c c c c c c}
  %\small{
  \hline\hline
  GRB/SN & z & $T_{90}^\dag$ & $E_{\gamma}$ & $L_{\gamma}$ & $M/M_{\bigodot}$ & $T_{90}/t_B$ & ref\\
  &  & [s] & [ergs] & [erg/s] & & & &\\
  \hline
  980425/1998bw & 0.0085 & 23 & $7\cdot10^{47}$ & $3\cdot10^{46}$ & 14 & 0.05 & 1,2\\
  031203/2003lw & 0.105 & 30 & $4\cdot10^{49}$ &$1.3\cdot10^{48}$ & 13 & 0.23 & 3,4\\
  051109B/ ? & 0.08 & 14 & $<1.3\cdot10^{49}$ &$<9\cdot10^{47}$ & $(15)^\ddag$ & $<0.09$ & 5\\
  060218/2006aj & 0.033 & 2000 & $6\cdot10^{49}$ &$3\cdot10^{46}$ & 3.3 & 7 & 6,7\\
  100316D/2010bh & 0.0593 & 1200 & $6\cdot10^{49}$ &$5\cdot10^{46}$ & 2.2 & 6 & 8,9\\
  %030329/2003dh & 0.168 & 25 & $8\cdot10^{51}$ & $4\cdot10^{50}$ & 8 & $2\cdot\eta^{-1/3}$ & 1,10\\
  \hline
  %}
   \multicolumn{8}{l}{\footnotesize $\dag$   redshift corrected values}\\
   \multicolumn{8}{l}{\footnotesize $\ddag$  assumed value due to lack of SN detection}\\
   \\
   \multicolumn{3}{l}{\footnotesize 1)\citet{Galama98}} & \multicolumn{2}{l}{\footnotesize 4)\citet{Mazzali06a}} &
   \multicolumn{2}{l}{\footnotesize 7)\citet{Mazzali06b}}\\
   \multicolumn{3}{l}{\footnotesize 2)\citet{Nakamura01}} & \multicolumn{2}{l}{\footnotesize 5)\citet{Troja06}} &
   \multicolumn{2}{l}{\footnotesize 8)\citet{Starling11}}\\
   \multicolumn{3}{l}{\footnotesize 3)\citet{Sazonov04}} & \multicolumn{2}{l}{\footnotesize 6)\citet{Soderberg06}} &
   \multicolumn{2}{l}{\footnotesize 9)\citet{Cano11}}\\
   %\multicolumn{3}{l}{\footnotesize 10)\citet{Mazzali03}}\\

 \end{tabular}
\end{table}

\section{The Origin of {\llGRBs}}

We have shown that it is quite unlikely that {\llGRBs} 
are  produced by  jets punching holes in their progenitors'
stellar envelopes. While we
cannot rule out that any
specific {\llGRB} was generated like that, the chances that the
group as a whole operates in this manner are small. Still, both
{\llGRBs} and LGRBs are accompanied by a rare type of SNe,
suggesting a strong connection between the two phenomena. Moreover,
late observations of {\llGRBs} accompanied SNe suggest that  their
progenitors harbor central engines \citep{Li99,Soderberg06}.

The two concepts can be reconciled if  {\llGRBs}'  jets simply
fail to breakout from their progenitors. A "failed jet" dissipates all
its energy into the surrounding cocoon and drives its expansion. As
the cocoon reaches the edge of the star its forward shock may become
mildly or even ultra relativistic emitting the observed $\gamma$-rays
when it breaks out. This idea that {\llGRBs} arise from
shock breakouts is not new. It was suggested shortly
following the observations of GRB980425/SN1998bw
\citep{Kulkarni98,MacFdyen01,Tan01}. It drew much more attention
following the observation of additional {\llGRBs} with similar
properties and especially with the observation of a thermal
component in the spectrum of {\llGRB} 060218
\citep{Campana06,Wang07,Waxman07}. Yet, it was hard to explain how
shock breakout releases enough energy in the form of $\gamma$-rays.
\cite{Katz10} realized that the deviation of the breakout
radiation from thermal equilibrium provides a natural explanation to
the observed $\gamma$-rays. More recently, \cite{Nakar11} calculated
the emission from mildly and ultra-relativistic shock breakouts,
including the post breakout dynamics and gas-radiation coupling.
They find that the total energy, spectral peak and duration of all
{\llGRBs} can be well explained  by relativistic shock
breakouts. Moreover, they find that such breakouts must satisfy a
specific relation between the observed total energy, spectral peak
and duration, and that all {\llGRBs} satisfy this relation. These
results lend a strong support to the idea that {\llGRBs} are
relativistic shock breakouts. From a historical point of view this
understanding 
closes the loop with Colgate's
(1968)  original idea, that preceded the detection of 
GRBs,
that  a SN shock breakout will produce a GRB.

In  a relativistic shock breakout the
burst duration is set by the properties of the shock and the
envelope at the edge of the star and not by the activity time of the
engine. Therefore, the engine may operate for only a short 
time, and still generate a very long burst like that of {\llGRBs}
060218 and 100316D. Thus, both  longer and  shorter duration
{\llGRBs} may be generated by this mechanism.

\section{Conclusions}

The activity of the
central engine, within the Collapsar model,
is independent of the breakout time of the jet.
This implies that the duration of the majority of bursts should  be
comparable to, or longer than, their jet breakout time
and that for $T_{90}\ll t_B$ 
the bursts' duration distribution should be
constant, independent of $T_{90}$. These two predictions are satisfied by
LGRBs, providing another support to the Collapsar model \citep{B11b}. As expected, they are
not satisfied by SGRBs that are not produced by Collapsars.

Like SGRBs, the observed distribution of {\llGRBs}  is inconsistent
with these  two predictions of the collapsar model. A  large
fraction of {\llGRBs} has $T_{90}/t_B\ll 1$.  The probability that
the observed {\llGRBs} $T_{90}/t_B$ distribution is consistent with
the LGRBs distribution is smaller than $5\%$. Similarly, the
{\llGRBs} $T_{90}$
distribution  (for $T_{90}\ll t_B$) is not flat 
at a confidence level  $>97\%$.

Taken together with their peculiar $\gamma$-ray emission, our overall conclusion %from the two tests
is that {\llGRBs} are
very unlikely to be produced by Collapsars 
like LGRBs. This can be
reconciled with the fact that {\llGRBs} are accompanied by ``engine
driven" SNe, if {\llGRBs} are produced by ``failed jets" that don't
break out of their progenitors. These ``failed" jets  deposit their energy into the
stellar envelopes and the $\gamma$-ray emission  arises from   shock breakout \citep{Kulkarni98,MacFdyen01,Tan01,Campana06,Wang07,Waxman07,Katz10,Nakar11}.
Our analysis doesn't prove this model but it
strongly disfavors its major 
competitor, any
variant on the Collapsar model.

The conclusion that the {\llGRBs} originate from "failed jets"
rather then successful ones has some interesting implications. In
particular the high rate of {\llGRBs} implies that jets which are
generated in SNe have a higher chance of remaining buried than to break
out. Accordingly 
most  SNe  engines can   generate jets
that produce 
 {\llGRB} and only a few are powerful enough to
produce jets that break out 
and produce  LGRBs \citep{Mazzali08}.

We thank to Re'em Sari for helpful discussions.
The research was supported by the Israel Center for Excellence for High Energy Astrophysics and
by an ERC advanced research grant (OB and TP) and by by the Israel Science Foundation
(grant No.\ 174/08) and by an EU International Reintegration Grant (EN).


\begin{thebibliography}{}

\baselineskip=15.8pt
%ApJ, 716, 781

%\bibitem[\protect\citeauthoryear{Bloom, Kulkarni, \& Djorgovski}{2002}]
%{Bloom02} Bloom J.~S., Kulkarni S.~R., Djorgovski S.~G.,
%2002, AJ, 123, 1111

%\bibitem[\protect\citeauthoryear{Bromberg \& Levinson}{2007}]{BL07} Bromberg O.,
%\& Levinson A., 2007, ApJ, 671, 678.

\bibitem[\protect\citeauthoryear{Bromberg et al.}{2011a}]{B11} Bromberg O.,
Nakar E., Piran T. \& Sari R., 2011, Submitted to ApJ (B11).

\bibitem[\protect\citeauthoryear{Bromberg et al.}{2011b}]{B11b} Bromberg O.,
Nakar E., Piran T. \& Sari R., 2011, In prep..


\bibitem[\protect\citeauthoryear{Campana et al.}{2006}]{Campana06}
Campana S., et al., 2006, Natur, 442, 1008

\bibitem[\protect\citeauthoryear{Cano et al.}{2011}]{Cano11}
Cano Z., et al., 2011, arXiv, arXiv:1104.5141

%\bibitem[\protect\citeauthoryear{Christensen, Hjorth,\& Gorosabel}{2004}]
%{Christensen04} Christensen L., Hjorth J.,
%Gorosabel J., 2004, A\&A, 425, 913

\bibitem[\protect\citeauthoryear{Cobb et al.}{2006}]{Cobb06}
Cobb B.~E., Bailyn C.~D., van Dokkum P.~G., Natarajan P., 2006, ApJ, 645,
L113

\bibitem[Colgate(1968)]{Cogate68} Colgate, S.~A.\ 1968, Canadian
Journal of Physics, 46, 476

\bibitem[\protect\citeauthoryear{Coward}{2005}]{Coward05}
Coward D.~M., 2005, MNRAS, 360, L77

\bibitem[\protect\citeauthoryear{Crowther}{2007}]{Crowther07}
Crowther P.~A., 2007, ARA\&A, 45, 177

\bibitem[\protect\citeauthoryear{Daigne \& Mochkovitch}{2007}]{Daigne07}
Daigne F., Mochkovitch R., 2007, A\&A, 465, 1

%\bibitem[\protect\citeauthoryear{Eichler \& Levinson}{1999}]{Eichler99}
%Eichler D., Levinson A., 1999, ApJ, 521, L117

\bibitem[\protect\citeauthoryear{Fan et al.}{2011}]{Fan11}
Fan Y.-Z., Zhang B.-B., Xu D., Liang E.-W., Zhang B., 2011, ApJ, 726, 32

%\bibitem[\protect\citeauthoryear{Fruchter et al.}{2006}]{Fruchter06}
%Fruchter A.~S., et al., 2006, Natur, 441, 463

\bibitem[\protect\citeauthoryear{Galama et al.}{1998}]{Galama98}
Galama T.~J., et al., 1998, ApJ, 497, L13

%\bibitem[\protect\citeauthoryear{Granot et al.}{2002}]{Granot02}
%Granot J., Panaitescu A., Kumar P., Woosley S.~E., 2002, ApJ, 570, L61

\bibitem[\protect\citeauthoryear{Guetta \& Della Valle}{2007}]{Guetta07}
Guetta D., Della Valle M., 2007, ApJ, 657, L73

\bibitem[\protect\citeauthoryear{Hjorth \& Bloom}{2011}]{Hjorth11}
Hjorth J., Bloom J.~S., 2011, arXiv, arXiv:1104.2274

\bibitem[\protect\citeauthoryear{Hullinger et al.}{2005}]{Hullinger05}
Hullinger D., et al., 2005, GCN, 4237, 1

%\bibitem[\protect\citeauthoryear{Kaneko et al.}{2007}]{Kaneko07}
%Kaneko Y., et al., 2007, ApJ, 654, 385

\bibitem[\protect\citeauthoryear{Katz, Budnik, \& Waxman}{2010}]{Katz10}
Katz B., Budnik R., Waxman E., 2010,

%\bibitem[Klebesadel et al.(1973)]{Klebesadel73} Klebesadel, R.~W.,
%Strong, I.~B., \& Olson, R.~A.\ 1973, \apjl, 182, L85

\bibitem[\protect\citeauthoryear{Kulkarni et al.}{1998}]{Kulkarni98}
Kulkarni S.~R., et al., 1998, Natur, 395, 663

%\bibitem[\protect\citeauthoryear{Lazzati, Morsony, \& Begelman}{2009}]{Lazzati09}
%Lazzati D., Morsony B.~J., Begelman M.~C., 2009, ApJ, 700, L47

%\bibitem[\protect\citeauthoryear{Le Floc'h et al.}{2003}]{Le Floc'h03}
%Le Floc'h E., et al., 2003, A\&A, 400, 499

\bibitem[\protect\citeauthoryear{Li \& Chevalier}{1999}]{Li99}
Li Z.-Y., Chevalier R.~A., 1999, ApJ, 526, 716

\bibitem[\protect\citeauthoryear{Liang et al.}{2007}]{Liang07}
Liang E., Zhang B., Virgili F., Dai Z.~G., 2007, ApJ, 662, 1111

\bibitem[\protect\citeauthoryear{MacFadyen\& Woosley}{1999}]{MacFadyen99}
MacFadyen A.~I., Woosley S.~E., 1999, ApJ, 524, 262

\bibitem[\protect\citeauthoryear{MacFadyen, Woosley, \& Heger}{2001}]{MacFdyen01}
MacFadyen A.~I., Woosley S.~E., Heger A., 2001, ApJ, 550, 410

%\bibitem[\protect\citeauthoryear{Matzner}{2003}]{Matzner03} Matzner C.~D.,
%2003, MNRAS, 345, 575

\bibitem[\protect\citeauthoryear{Matzner \& McKee}{1999}]{MatzMck99}
Matzner C.~D., McKee C.~F., 1999, ApJ, 510, 379


\bibitem[\protect\citeauthoryear{Mazzali et al.}{2003}]{Mazzali03}
Mazzali P.~A., et al., 2003, ApJ, 599, L95

\bibitem[\protect\citeauthoryear{Mazzali et al.}{2006a}]{Mazzali06a}
Mazzali P.~A., et al., 2006a, ApJ, 645, 1323

\bibitem[\protect\citeauthoryear{Mazzali et al.}{2006b}]{Mazzali06b}
Mazzali P.~A., et al., 2006b, Natur, 442, 1018

%\bibitem[\protect\citeauthoryear{Nakamura}{1998}]{Nakamura98}
%Nakamura T., 1998, PThPh, 100, 921

%\bibitem[\protect\citeauthoryear{Nakamura}{1999}]{Nakamura99}
%Nakamura T., 1999, ApJ, 522, L101
\bibitem[\protect\citeauthoryear{Mazzali et al.}{2008}]{Mazzali08}
Mazzali P.~A., et al., 2008, Sci, 321, 1185


\bibitem[\protect\citeauthoryear{Nakamura et al.}{2001}]{Nakamura01}
Nakamura T., Mazzali P.~A., Nomoto K., Iwamoto K., 2001, ApJ, 550, 991

\bibitem[\protect\citeauthoryear{Nakar \& Sari}{2011}]{Nakar11}
Nakar E., Sari R., 2011, submitted to ApJ.

\bibitem[\protect\citeauthoryear{Paczynski}{1998}]{Paczynski98}
Paczynski B., 1998, ApJ, 494, L45

\bibitem[\protect\citeauthoryear{Pian et al.}{2006}]{Pian06}
Pian E., et al., 2006, Natur, 442, 1011

%\bibitem[\protect\citeauthoryear{Ramirez-Ruiz et al.}{2005}]{Ramirez05}
%Ramirez-Ruiz E., Granot J., Kouveliotou
%C., Woosley S.~E., Patel S.~K., Mazzali P.~A., 2005, ApJ, 625, L91

\bibitem[\protect\citeauthoryear{Sazonov, Lutovinov, \& Sunyaev}{2004}]{Sazonov04}
Sazonov S.~Y., Lutovinov A.~A., Sunyaev R.~A., 2004, Natur, 430, 646

\bibitem[\protect\citeauthoryear{Soderberg, Frail, \& Wieringa}{2004}]{Soderberg04a}
Soderberg A.~M., Frail D.~A., Wieringa M.~H., 2004a, ApJ, 607, L13

%\bibitem[\protect\citeauthoryear{Soderberg et al.}{2004b}]{Soderberg04b}
%Soderberg A.~M., et al., 2004b, Natur, 430, 648

\bibitem[\protect\citeauthoryear{Soderberg et al.}{2006a}]{Soderberg06}
Soderberg A.~M., et al., 2006a, Natur, 442,1014

\bibitem[\protect\citeauthoryear{Soderberg et al.}{2006b}]{Soderberg06b}
Soderberg A.~M., Nakar E., Berger E., Kulkarni S.~R., 2006b, ApJ, 638, 930


%\bibitem[\protect\citeauthoryear{Sollerman et al.}{2006}]{Sollerman06}
%Sollerman J., et al., 2006, A\&A, 454, 503

%\bibitem[\protect\citeauthoryear{Sparre et al.}{2011}]{Sparre11}
%Sparre M., et al., 2011, arXiv, arXiv:1105.0422

%\bibitem[\protect\citeauthoryear{Stanek et al.}{2003}]{Stanek03}
%Stanek K.~Z., et al., 2003, ApJ, 591, L17

\bibitem[\protect\citeauthoryear{Starling et al.}{2011}]{Starling11}
Starling R.~L.~C., et al., 2011, MNRAS,
411, 2792

\bibitem[\protect\citeauthoryear{Tan, Matzner, \& McKee}{2001}]{Tan01}
Tan J.~C., Matzner C.~D., McKee C.~F., 2001, ApJ, 551, 946

\bibitem[\protect\citeauthoryear{Troja et al.}{2006}]{Troja06}
Troja E., Cusumano G., Laparola V., Mangano V., Mineo T., 2006, NCimB, 121,
1599

\bibitem[\protect\citeauthoryear{Wanderman \& Piran}{2010}]{Wanderman10}
Wanderman D., Piran T., 2010, MNRAS, 406, 1944

\bibitem[\protect\citeauthoryear{Wang et al.}{2007}]{Wang07}
Wang X.-Y., Li Z., Waxman E., M{\'e}sz{\'a}ros P., 2007, ApJ, 664, 1026

%\bibitem[\protect\citeauthoryear{Watson et al.}{2004}]{Watson04}
%Watson D., et al., 2004, ApJ, 605, L101

\bibitem[\protect\citeauthoryear{Waxman}{2004}]{Waxman04}
Waxman E., 2004, ApJ, 602, 886

\bibitem[\protect\citeauthoryear{Waxman, M{\'e}sz{\'a}ros, \& Campana}{2007}]{Waxman07}
Waxman E., M{\'e}sz{\'a}ros P., Campana S., 2007, ApJ, 667, 351


%\bibitem[\protect\citeauthoryear{Woods \& Loeb}{1999}]{Woods99}
%Woods E., Loeb A., 1999, ApJ, 523, 187

\bibitem[\protect\citeauthoryear{Woosley \& Bloom}{2006}]{Woosley06}
Woosley S.~E., Bloom J.~S., 2006, ARA\&A, 44, 507


%\bibitem[\protect\citeauthoryear{Woosley, Eastman, \& Schmidt}{1999}]{Woosley99}
%Woosley S.~E., Eastman R.~G., Schmidt B.~P., 1999, ApJ, 516, 788




\end{thebibliography}
\end{document}